\documentclass[aps,pre,twocolumn,keywords,showpacs]{revtex4}
\usepackage{graphicx}
\usepackage{amsmath}
\begin{document}


\title{Chemically driven electron tunnelling  pumps}
\author{Igor Goychuk}
\affiliation{Institut f\"ur Physik,
Universit\"at Augsburg, Universit\"atsstra\ss e 1, D-86135 Augsburg,
Germany}


\begin{abstract}
The simplest mechanism for molecular electron pumps is discussed which
is based on nonadiabatic electron tunnelling and nonequilibrium conformational
fluctuations. Such fluctuations can be induced, e.g. by 
random binding of negatively charged ATP molecules to the electron-transferring
molecular complex,
their subsequent hydrolysis and the products dissociation. The pumping
rate can be controlled by the ATP concentration in solution. Depending on
the model parameters there may exist a critical ATP concentration for the pump
to function. 
Alternatively,
nonequilibrium fluctuations can be induced 
by externally applied
stochastic electric fields. For realistically chosen parameters,  
the mechanism is shown to be robust and highly efficient.
\end{abstract}


\maketitle

\section{Introduction}

Electron transfer lies at heart of all bioenergetic processes. 
The energy of photoexcited electronic states, or one released
in the oxidative breakdown of food molecules is used  in a
chain of electron-transfer reactions to create the  transmembrane
proton gradient storing ultimately the free energy ready to use  in
such "energetic" organels of biological cells as mitochondria and 
chloroplasts \cite{Mitchell1,Mitchell2,Alberts,NossalLecar,Nelson}.  
This electrochemical proton
gradient is used there by the ATP-syntase molecular complexes 
to synthesise the molecules of adenosintriphosphate 
(ATP) -- 
a free-energy ``currency''
utilised in the most biochemical cellular  processes which occur far
away from the thermodynamic equilibrium under the living cell conditions.
It is well known that the ATP-syntase can work also in reverse using the
energy of ATP hydrolysis to restore the proton gradient \cite{Alberts}.  
The existence of the reverse
electron transfer, where the energy of ATP hydrolysis, or the free energy
derived  from  electrochemical transmembrane gradient of a sort of
ions is used to energise the electrons, is coming gradually in the 
focus of attention \cite{Chance,Miki,Elbenti,Gemperli,Osyczka}. 
The proton pumping molecular complexes driven normally by the energy
released in the downhill electron transfer can work in reverse, pumping the
electrons uphill on the time scale of seconds and minutes \cite{Miki,Osyczka}.  
It might even
be the case that such reverse electron transfer evolutionary emerged 
earlier in archaebacteria  existing at
extremal environmental  conditions (i.e., during the most earliest
steps of the biological evolution).  
Moreover, the nitrogen fixation, which is realized by the nitrogenase
protein  complexes,  utilises apparently  the energy released in
ATP hydrolysis \cite{Lanzilotta,Rees,Davidson,Kurnikov}.
These natural  molecular nanomachines  produce ammonia routinely, 
at normal conditions,  while the standard industrial technological
process requires large
pressures of about 150 atmospheres and temperatures in the range of 650-720 K. 

The electron transfer in  nitrogenase provides one of the key steps
in the overall reaction of ammonia synthesis. It is
realized through a long-distance (about 14 \AA)  nonadiabatic 
electron tunnelling
between two different metalloclusters situated in two different 
protein subunits. This process is gated by nonequilibrium  conformational
transitions of the whole protein complex  
due to ATP binding and hydrolysis, with two ATP molecules
hydrolysed per one  electron transferred.  How nitrogenase 
works remains still a
mystery, but a proper physical understanding gradually emerges \cite{Kurnikov}.
Such understanding is crucial not only for uncovering
the working principles of 
biological nanomachinery in general, but also for the molecular 
design of such and similar
molecular machines for  a future nanobiotechnological use.  It would allow
for an intelligent  parameter optimisation, performing  ultimately better than
nature. Below I consider a very simplified,  minimal theoretical model  for such
chemically driven electron tunnelling pumps which is based on the previous
treatments in \cite{Kurnikov,JCP95,PRE95,JCP97,PRE97}.

\section{Theoretical model}

\begin{figure}[ht]
\begin{center}
 \includegraphics[width=8cm]{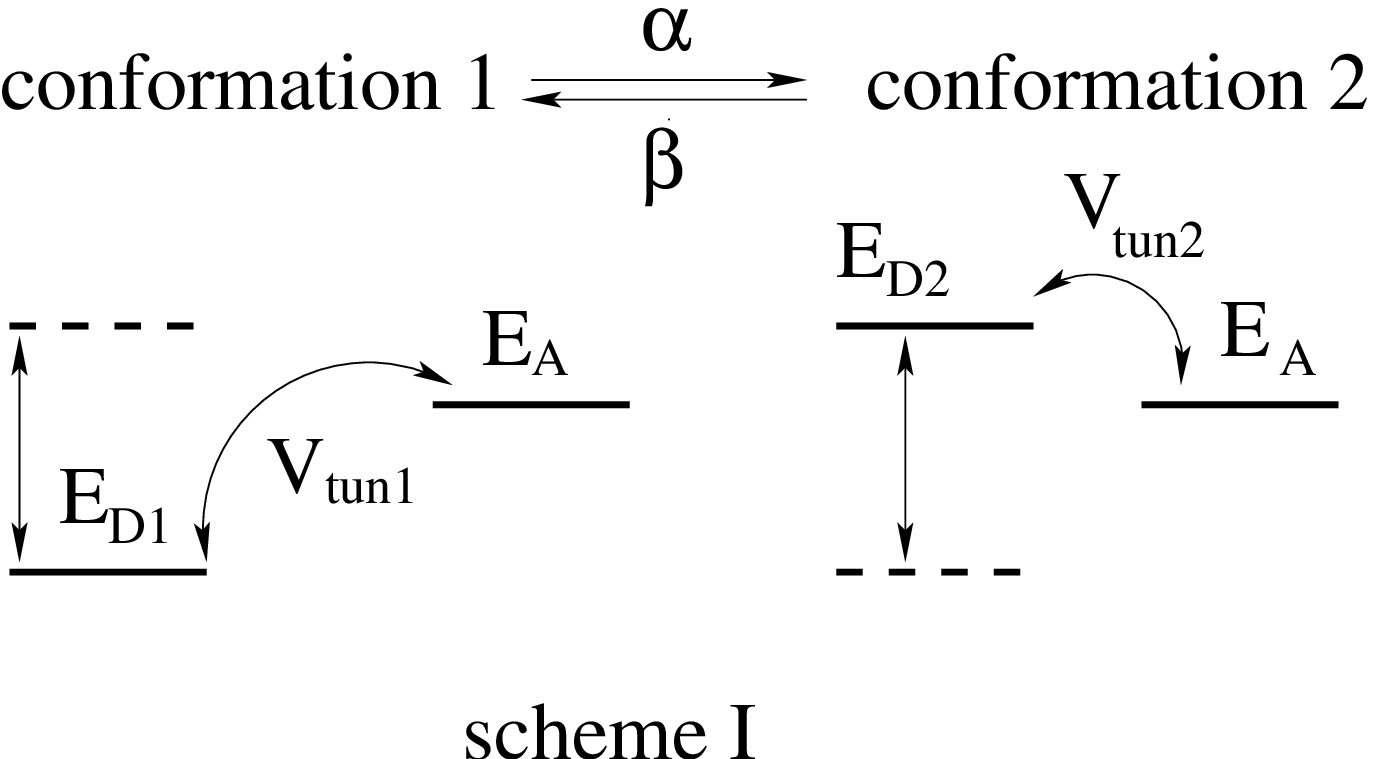}
\vfill
\vspace{0.5cm} 
 \includegraphics[width=8cm]{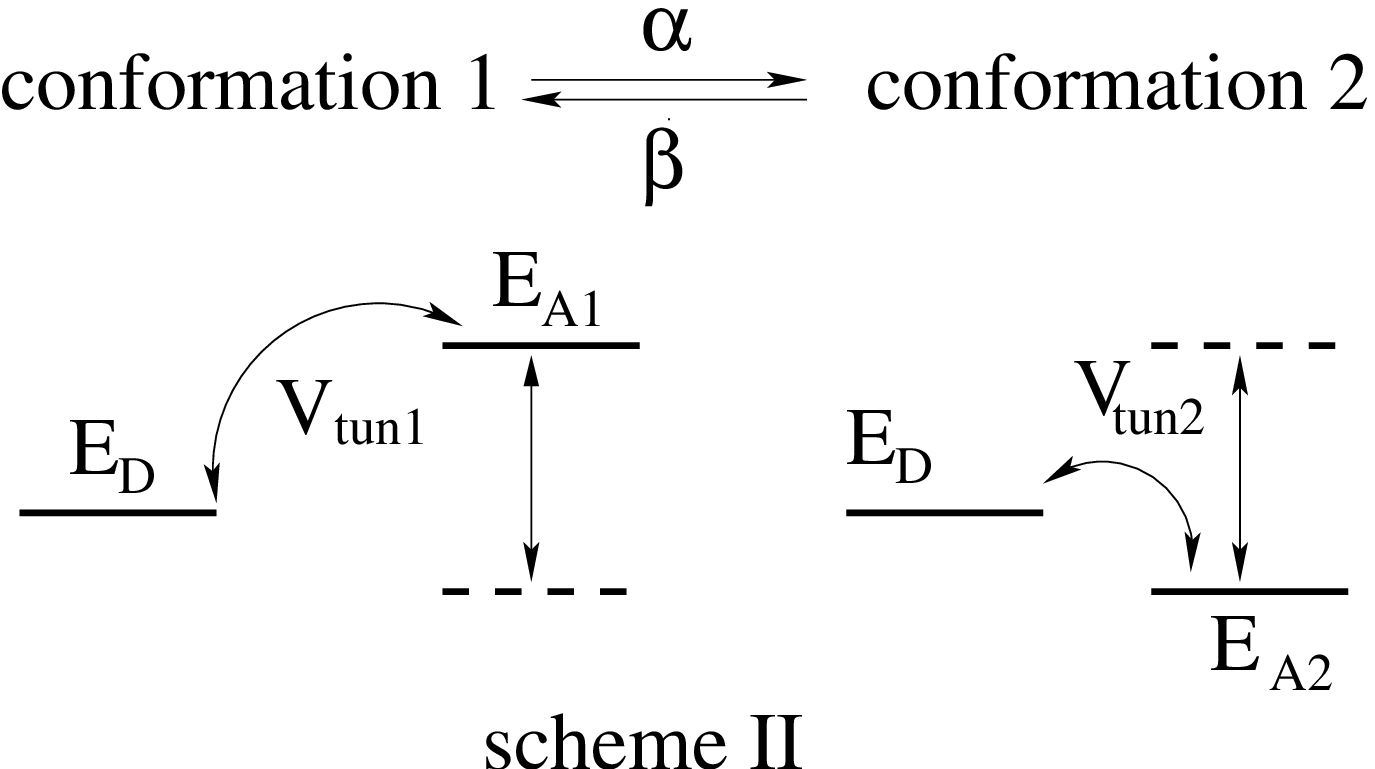}
\end{center}

 \caption{Electron pumping scenarios based on dissipative electron
 tunnelling and nonequilibrium
 conformational fluctuations. In the scheme I, the energy level of the 
 donor state fluctuates in time. This is contrasted with the scheme II,
 where the acceptor level fluctuates in time. 
 For dissipative tunnelling to occur from the donor site 
 to the acceptor site 
 of electron localisation,
 either the donor state should gain temporally in energy, or the
 acceptor cite should temporally lose in energy (conformation 2). 
 A combination of
 two scenarios is also possible.
 The condition $V_{\rm tun1} \ll V_{\rm tun2}$ is required
 for the pumping mechanism to work robustly. It is necessary to block
 the reverse acceptor-to-donor electron transfer in the conformation 1. 
 Temporal lifting of the 
 electron energy is possible, e.g. due to binding a negatively charged
 ATP molecule nearby the corresponding site of localisation. The
 tunnelling coupling can be exponentially reduced, e.g. due to increase
 of the tunnelling distance, or due to disruption of the 
 tunnelling pathway. The latter one can be  induced also 
 by reorientation of a 
 bridging molecular group. \label{Fig1}}
 \end{figure}

\subsection{Nonadiabatic electron transfer}
 
Let us start from nonadiabatic electron tunnelling coupled to molecular
vibrational modes considered within a standard two-state, donor acceptor model
(cf. one ``frozen'' conformation in Fig.\ref{Fig1}). 
This tunnelling
can be described, e.g., within a spin-boson like model captured by the
following Hamiltonian 
\cite{Garg,Onuchic,Schulten,Leggett,May,PetrovBook,AdvPhys05}
\begin{eqnarray}\label{Hamiltonian}
\hat H=E_D|D\rangle \langle D|+ E_A|A\rangle \langle A|+
V_{\rm tun}(|D\rangle \langle A|+|A\rangle \langle D|)\nonumber\\
+\frac{1}{2}(|D\rangle \langle D|-|A\rangle \langle A|)
\sum_j \kappa_j(\hat b^{\dagger}_j+\hat b_j) \nonumber \\
+ \sum_{j}\hbar\omega_j(\hat b^{\dagger}_j\hat b_j+1/2).
\end{eqnarray}
Wherein, $|D\rangle$ and $|A\rangle$, are the localised electronic states
with the energies $E_D$ and $E_A$, correspondingly;  $V_{\rm tun}$ denotes the 
effective electron tunnelling matrix
element which incorporates the intervening medium influence  
(a superexchange tunnelling mechanism is assumed
\cite{McConnell,KPU78,Beratan87,Skourtis99,Gray05}).  
The coupling of electron tunnelling to the molecular
vibrational modes $\omega_j$ is characterised by the coupling constants 
$\kappa_j$ and the corresponding spectral density $J(\omega)=(2\pi/\hbar^2)
\sum_j\kappa_j^2\delta(\omega-\omega_j)$. This coupling modulates 
the energy difference between the donor and acceptor states. The corresponding
fluctuations are described by (quantum) random force 
 $\hat\xi(t)=\sum_{j}\kappa_{j}(b^{\dagger}_{j} e^{i\omega_{j}t}+b_{j}
 e^{-i\omega_{j}t})$ with thermally equilibrium autocorrelation 
 function $\langle\hat{\xi}(t)\hat{\xi}(0)\rangle_{T}=
 \frac{\hbar^2}{2\pi}\int\limits_{0}^{\infty} J(\omega)[
 \coth(\hbar\omega/2k_BT) \cos(\omega t)-i\sin\omega t ] d\omega$.
The medium's reorganisation energy $\lambda=\hbar\int_{0}^{\infty}d\omega 
J(\omega)/(2\pi\omega)$ serves is an integral 
characteristics of this coupling. Another important medium's characteristic
is the upper frequency $\omega_c$ 
of low-frequency molecular vibrations, or solvent modes 
coupled to the electron transfer (ET), i.e.
$J(\omega)=0$ for $\omega\gg \omega_c$. 
 For $V_{\rm tun}\ll \lambda, \sqrt{k_BT\hbar\omega_c}$, the transfer
 kinetics occurs in the nonadiabatic tunnelling regime
described by the quantum master equations of the Pauli type for the populations
of donor and acceptor states 
\cite{Garg,Onuchic,Schulten,Leggett,May,PetrovBook,AdvPhys05}, 
\begin{eqnarray}\label{Pauli}
\dot p_D(t)& = &-k_{\rm f}\;p_D+k_{\rm b}\;p_A, \nonumber \\
\dot p_A(t)& = &-k_{\rm b}\;p_A +k_{\rm f}\;p_D \;, 
\end{eqnarray}
with the forward, donor-to-acceptor rate given 
by the quantum Golden Rule expression     
\begin{eqnarray}\label{rate}
 k_{\rm f}=\frac{2V_{\rm tun}^2}{\hbar^2}\int_{0}^{\infty}
 d\tau\exp[-Q^{\prime}(\tau)]\cos\left[Q^{\prime\prime}(\tau) -
 \epsilon\tau/\hbar\right] \;,
 \end{eqnarray}
where $\epsilon=E_D-E_A$ is the difference of free energies (or the
thermodynamic driving
force $-\Delta G$ in chemical notations, $\epsilon=-\Delta G$). Furthermore,
the functions $Q^{\prime}(t)$
and $Q^{\prime\prime}(t)$ in (\ref{rate}) denote the real and
imaginary parts of the doubly-integrated bath autocorrelation function
with the reorganisation energy contribution added,
 \begin{equation}\label{Q-fun}
 Q(t)= \frac{1}{\hbar^2}
 \int_{0}^{t}dt_1\int_{0}^{t_1}
 \langle \hat \xi(t_2)\hat \xi(0)\rangle_{T} dt_2+i\lambda t/\hbar \;.
 \end{equation}
The backward rate  $k_{\rm b}$ satisfies the Boltzmann relation
\begin{equation}\label{Boltzmann}
k_{\rm b}=k_{\rm f}\exp(-\epsilon/k_BT)
\end{equation}
for any $J(\omega)$ \cite{Leggett}.  This yields equilibrium Boltzmann-Gibbs distribution 
$p_D(\infty)/p_A(\infty)=\exp(-\epsilon/k_BT)$ at 
temperature $T$, consistently with the condition of detailed balance, 
$p_D(\infty)k_{\rm f}=p_A(\infty)k_{\rm b}$. It must be stressed, 
however, that 
the existence of the Boltzmann
relations for rates like one in Eq. (\ref{Boltzmann}) at constant 
$\epsilon$ and $V_{\rm tun}$ 
does not guarantee, generally, the detailed balance and the 
Boltzmann-Gibbs equilibrium distribution when these parameters
explicitly fluctuate in time \cite{AdvPhys05}, or when a stationary flux
is present (given open, or cyclic boundary conditions even at constant
rates). This is true
even if the rates follow adiabatically
to the instant values of the energy levels and the tunnelling
coupling so that the condition (\ref{Boltzmann}) is valid at any
instant of time.

Independently of other details of electron-vibrational coupling, 
the quantum Golden Rule rate acquires  in the 
high-temperature limit $k_BT \gg\hbar\omega_c$ and the quasi-static 
approximation of electron energy fluctuations, 
$\langle \hat \xi(t)\hat \xi(0)\rangle_{T}\approx 
\langle \hat \xi^2(0)\rangle_{T}\approx 2\lambda k_BT$, 
the semiclassical Marcus-Levich-Dogonadze form \cite{Marcus,Levich}
\begin{eqnarray}\label{Marcus}
k_{\rm f}=\frac{2\pi}{\hbar} \frac{V_{\rm tun}^2}{
\sqrt{4\pi\lambda k_BT}}\exp[-(\epsilon
  -\lambda)^2/(4\lambda k_B T)]\; .
 \end{eqnarray}
Such a universality explains the widespread use of Eq. (\ref{Marcus}) 
in interpretation of experimental data. 
It involves (apart from temperature) three parameters only:
the free energy difference $\epsilon$, the electron tunnelling 
coupling $V_{\rm tun}$ and the reorganisation energy $\lambda$. 
When a coupling to high-frequency, essentially quantum modes is present, 
the expression (\ref{Marcus}) can be readily  
generalised accordingly \cite{Jortner,BixonJortner}. 
This presents one of the basic
theories of electron transfer in molecular systems 
\cite{BixonJortner,May,PetrovBook}. Of course, it is not truly
universal, but it
does provide a milestone and the simplest theoretical framework
of practical relevance. 

\subsection{Nonequilibrium conformational fluctuations}

Furthermore, let us assume that the electron-transferring protein complex
(or its corresponding molecular subunit) can be in either of 
two conformations
depending on binding a ligand, say ATP molecule (cf. Fig. \ref{Fig1}). 
This assumption 
is quite in spirit of the Monod-Wyman-Changeux model of allosteric
enzymes \cite{Monod,KeenerSneyd,Changeux}. These two conformations
possess very different $V_{\rm tun}$ (distance between the donor and acceptor
sites is changed, or some bridging molecular group changes its orientation
interrupting, or vice versa, establishing thereby the electron-tunnelling 
pathway). These conformations can correspond also to very different
energy differences  between the localised 
electron levels, e.g. the energy of the donor or acceptor
state is changed (ATP and the hydrolysis 
products, ADP and the phosphate group $\rm P_i$, are all charged and 
the electrostatic effects are of utmost importance here \cite{Kurnikov}). 
The attachment/detachment
of ligand is a random process and both $\epsilon(t)=E_D(t)-E_A(t)$ 
and $V_{\rm tun}(t)$ in
Eq. (\ref{Hamiltonian}) become stochastic functions of time.
Alternatively, a strong two-state stochastic electric field can be externally
applied to drive the electron transfer process. 
This is the starting point
of the stochastically driven spin-boson model of 
Refs. \cite{JCP95,PRE95,JCP97,
PRE97,AdvPhys05}. It must be noted that modelling of the equilibrium 
conformational fluctuations in such a way should be considered with a great
care \cite{AdvPhys05} (see also below), but the approach suits well to model 
non-equilibrium fluctuations
like those considered \cite{AdvPhys05}. Within the approximations
leading to Eqs. (\ref{Pauli})-(\ref{Q-fun}), the quantum rates entering
equations (\ref{Pauli}) become stochastic {\it functionals} 
of $V_{\rm tun}(t)$ and $\epsilon(t)$. Namely, the term $\epsilon\tau$
in Eq. (\ref{rate}) is replaced by $\int_{t-\tau}^t\epsilon(t')dt'$
and instead of $V_{\rm tun}^2$ in the front of integral there appears
$V_{\rm tun}(t)V_{\rm tun}(t-\tau)$ in the integrand 
\cite{JCP95,PRE95}. However, if $\epsilon(t)$ and $V_{\rm tun}(t)$ 
fluctuate slow on the time scale of $Q(t)$, one can use 
an {\it adiabatic driving} approximation resulting in fluctuating rates 
following to the
instantaneous values of $\epsilon(t)$ and $V_{\rm tun}(t)$. This 
approximation is reasonable as a simple starting point for modelling
 and it can be justified
in many cases. For these reasons, it is used  below. 
The discussed adiabatic assumption 
means that after every conformational jump 
the {\it vibrational} relaxation to the new equilibrium of the
vibrational degrees of freedom occurs very fast as compare
with the mean duration of time spent in the corresponding conformation.
Otherwise, the adiabatic driving approximation cannot be justified and
the theory becomes essentially more intricate
\cite{JCP95,PRE95,JCP97,PRE97}. 

\begin{figure}[ht]
\begin{center}
 \includegraphics[width=8cm]{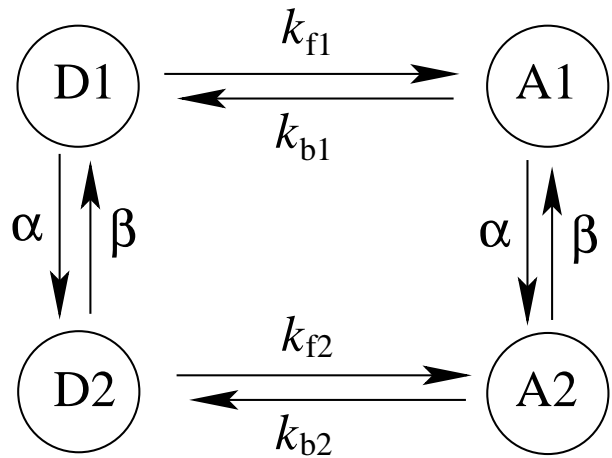}
 \end{center}
 \caption{Equivalent kinetic scheme corresponding to quantum kinetic 
 equations (\ref{Pauli}) with rates undergoing two-state
 Markovian fluctuations \label{Fig2}}
  \end{figure}

Furthermore, let us assume 
that the conformational fluctuations are Markovian and occur with 
the rates $\alpha$ and $\beta$
which do not
depend on where the electron is localised, but are controlled
by thermodynamically nonequilibrium concentrations of 
ATP, ADP, $\rm P_i$ in the solution 
(i.e. ATP is continuously supplied \cite{foot1}).
To be more concrete, let us assume that the conformational transition
``1$\to$ 2'' is caused by the ATP binding to the electron-transferring
molecular complex (scheme I). Then, the transition rate $\alpha$ should obviously
be proportional to the ATP concentration, [ATP], in the solution, i.e. 
$\alpha\propto \rm [ATP]$, since the binding frequency is proportional
to [ATP]. On the contrary, the rate $\beta$ of the 
conformational transition ``2$\to$1'' caused by the ATP hydrolysis
and the products dissociation should not depend on [ATP], but 
be rather determined by the activation barrier between two conformations
and the energy released by breaking the phosphate bond.    
Such nonequilibrium fluctuations fuelled by this, or another source
of chemical energy  can drive electron transfer (ET) uphill.
Alternatively, they can
be induced by an externally applied stochastic electric field 
\cite{Tsong,Westerhoff}.
It can either be directly coupled to the electron transfer 
\cite{PRE95,JCP97,PRE97}, or modulate the electron levels indirectly,
via the electroconformational coupling 
\cite{Tsong}.  
Then, 
within the discussed
approximations, the ET transfer kinetics is described 
by the kinetic equations (\ref{Pauli}) with time-dependent rates 
undergoing two-state Markovian fluctuations. Formally, this is  a typical
problem of dynamical disorder 
\cite{Zwanzig,KampenBook,Hoffman,Gehlen}.
It can equivalently be described by the 
four-state Markovian kinetic scheme depicted in Fig. \ref{Fig2}
\cite{foot2}.
Similar schemes are standard by considering the problem of 
free energy transduction in biology \cite{Hill1,Hill2,Hill3}.

Let us denote the population of the donor and acceptor states in the 
conformation $i=1,2$ as $p_{Di}$ and $p_{Ai}$, correspondingly. They 
obviously satisfy the master equations
\begin{eqnarray}\label{master-eqs}
\dot p_{D1}& = & -(k_{\rm f1}+\alpha) p_{D1}+\beta p_{D2}+
k_{\rm b1}p_{A1} \;, \nonumber\\
\dot p_{D2}& = & \alpha  p_{D1} -(\beta+k_{\rm f2}) p_{D2}+
k_{\rm b2}p_{A2} \;,\nonumber \\
\dot p_{A1}& = & k_{\rm f1} p_{D1} -(\alpha+k_{\rm b1}) p_{A1}+
\beta p_{A2}\;, \\
\dot p_{A2} & = & k_{\rm f2} p_{D2} +\alpha p_{A1}-
(\beta + k_{\rm b2})p_{A2}\;. \nonumber
\end{eqnarray}
These equations describe the process at thermodynamical equilibrium if
the overall stationary flux is absent, i.e. clockwise and 
counterclockwise fluxes are mutually compensated. This requires that 
the product of forward rates
along the cycle is equal to the product of backward rates, see e.g. in 
\cite{Hill2,Westerhoff2,Qian,Min}.
Otherwise, a nonequilibrium steady state (NESS) emerges
with a persistent flux present. In such a case,
one either requires a free energy supply to produce
the corresponding stochastic cyclic motion (uphill motion on an effective
free energy landscape), or 
this energy will be released (in the downhill motion). Let us consider
the situation where the averaged free energy bias $\overline{\epsilon}=
\epsilon_1p_1^{st}+\epsilon_2p_2^{st}$ is negative,  
$\overline{\epsilon}<0$ ($p_1^{st}$ and $p_2^{st}$
are the stationary probabilities of the corresponding conformations).
Then the cycling in the counterclockwise direction in Fig. 
\ref{Fig2}   is required to pump the electrons
against the averaged free energy bias $\overline{\epsilon}<0$ -- the
case of our interest here. If NESS corresponds to the clockwise
total probability flux in Fig. \ref{Fig2}, the roles of donor and 
acceptor states are interchanged 
and the conformational fluctuations can
be driven by the energy released in the downhill
electron transfer. These conformational fluctuations can in turn be coupled
to an ion flux 
to produce the uphill ion flow against the corresponding electrochemical 
gradient. 
This is the operating principle of the electron-driven proton pumps 
\cite{Alberts,Westerhoff2}. 
The details
are, of course, much more involved (a more complex, extended 
kinetic scheme is required to describe these
processes in a consistent manner) and still not completely
understood. The operating principle is, however, rather clear
due to (nonlinear)
nonequilibrium thermodynamics considerations. In the present context,
we combine them with a quantum treatment of the electron transfer
kinetics. 
 
The free energy $\Delta G_{\rm drive}$, which is required to drive 
one cycle on average in the counterclockwise direction in Fig. \ref{Fig2} 
is determined by the well-known condition \cite{Hill2}
\begin{eqnarray}
\Delta G_{\rm drive}=k_B T\ln \left
(\frac{k_{\rm f2}\beta k_{\rm b1}\alpha}{k_{\rm b2} \beta 
k_{\rm f1} \alpha } \right)\\ \nonumber
=k_B T\ln \left
(\frac{k_{\rm f2}k_{\rm b1}}{k_{\rm b2}  
k_{\rm f1}} \right)=\epsilon_2-\epsilon_1=E_{D2}-E_{D1}.
\end{eqnarray}
For the scheme II, $\Delta G_{\rm drive}=E_{A1}-E_{A2}$.
In the second line, we took Eq. (\ref{Boltzmann}) into account.
As discussed above,
this energy can be delivered, e.g., due to the ATP hydrolysis (breaking
the energy rich phosphate bond), 
or derived from any
other free energy source (the proton gradient, for example, via 
protonation/deprotonation of the molecular pump). 
Obviously, if only the tunnelling matrix element
is modulated by the conformational transitions, then no pumping is possible
and the described scheme agrees with the thermodynamic equilibrium, since
no overall flux is present. We return back to
the well-known problem of dynamical disorder in equilibrium systems 
and no more. To pump, one
has to modulate the difference of electron energy levels $\epsilon$ in time. 
With $V_{\rm tun}=const$ a pumping scenario
conditioned on the existence of inverted Marcus regime of electron transfer,
where the transfer rate decreases with the increase of the energy bias,
is possible. It was described in Ref. \cite{PRE97}. In this respect, it is
worth to notice that the very existence of the inverted ET regime 
presents a profoundly quantum-mechanical feature of nonadiabatic ET.
Therefore, such an electron pump would be essentially quantum-mechanical. 
However, such a pumping
scenario based on the sole modulation of $\epsilon(t)$ would 
be rather inefficient and too demanding for 
the system parameters in practice. 
Therefore, it is not likely to be used by nature. On the contrary,
a properly concerted modulation
of $\epsilon(t)$ and $V_{\rm tun}(t)$ can allow to pump highly 
efficiently. In essence, for this 
one has to ensure $V_{\rm tun1}\ll V_{\rm tun2}$ in Fig. \ref{Fig1}
and a proper
timing when ET kinetics is gated and locked to conformational
fluctuations. Two possible pumping scenarios are depicted in Fig. \ref{Fig1}.
In the scheme I,  the donor energy level
is lifted upon binding negatively charged ATP molecule(s).
Alternatively, one can modulate the acceptor energy level in time, 
scheme II in Fig. \ref{Fig1}.  In the reality, a combination of both
possibilities can take place. For example, in the case of nitrogenase
the donor level is lifted by 300 meV and the acceptor level increases
simultaneously by 100 meV,
with the total increase of the driving energy bias by 200 meV 
\cite{Kurnikov}. This compares well with the energy release from the 
hydrolysis of one ATP molecule which is about $0.3-0.5$ eV 
under the living cell conditions.
This basic pumping mechanism will be detailed and quantified below.

\subsection{Solution of the model}

How to proceed further
is  standard and well-known \cite{KampenBook}. 
The solution of the master equations (\ref{master-eqs}) 
with the initial conditions $p_{j'}(0)=1$, where $j'=D1,D2,A1,A2$ yields
the corresponding conditional probabilities of the state $j$, $P_{jj'}(t)$. 
We are interested in several quantities, such as (i) the asymptotic population
of the donor state $\langle p_D(\infty)\rangle$; (ii) the 
time course of the donor state relaxation $\langle p_D(t)\rangle$, provided
that the electron was initially prepared in the donor state; (iii) the
distribution of the first arrival times, $\psi_D(\tau)$, at the acceptor 
state and the corresponding mean forward transfer time $\langle \tau_f\rangle:=
\int_0^{\infty}\tau\psi_D(\tau)d\tau=\int_0^{\infty}\Phi_D(\tau)d\tau$,
where $\Phi_D(\tau)=\int_{\tau}^{\infty}\psi_D(\tau)d\tau$ is the corresponding
survival probability. Given the conditional probabilities $P_{jj'}(t)$, the
contracted probability of states $D1$ and $D2$ with the electron being localised initially
on the donor site is
\begin{eqnarray} \label{pd1}
\langle p_{D1}(t)\rangle=P_{D1D1}(t)p_{1}^{st}+P_{D1D2}(t)p_{2}^{st}
\end{eqnarray}
and
\begin{eqnarray} \label{pd2}
\langle p_{D2}(t)\rangle=P_{D2D1}(t)p_{1}^{st}+P_{D2D2}(t)p_{2}^{st},
\end{eqnarray}
correspondingly. In the above equations, it is tacitly assumed that
the donor site has the same affinity to the transferring excess electron
in the both protein conformations and these conformations are met with the
stationary probabilities $p_{1}^{st}=\beta/(\alpha+\beta)$ and 
$p_{2}^{st}=\alpha/(\alpha+\beta)$. The former assumption is trivially valid
for the scheme II. However, for the scheme I it might be the case
that the electron affinity to D2 state is much smaller than to D1, i.e. the 
protein binds the transferring electron with a much higher probability in the
 first conformation (when no negatively charged ATP is bound nearby the donor state).
  In such a
situation (which is not considered here for the sake of simplicity 
and analytical tractability of the results), 
one should put $p_{1}^{st}\to 1$ and $p_{2}^{st}\to 0$ in Eqs.
(\ref{pd1}) and (\ref{pd2}). In any case, the averaged population 
of the donor state
is $\langle p_{D}(t)\rangle=\langle p_{D1}(t)\rangle+\langle p_{D2}(t)\rangle$.

The formal solution can be found most conveniently using the
Laplace-transform method. After some lengthy algebra we obtain (assuming  equal
electron binding affinities of the D1 and D2 states): 
\begin{eqnarray}\label{pd}
\langle \tilde p_{D}(s)\rangle=\frac{1}{s}\frac{\tilde A(s)}{\tilde B(s)},
\end{eqnarray}
where 
\begin{eqnarray}
\tilde A(s) &= &s^2 +[\nu+p_2^{st}(k_1+k_{\rm b2})\nonumber \\
&& +p_1^{st}(k_2+k_{\rm b1})]s \\
& & + p_2^{st}k_{\rm b2}(k_1+\nu)+p_1^{st}k_{\rm b1}(k_2+\nu),\nonumber\\
\tilde B(s) & = &s^2+(\nu+k_1+k_2)s +k_1k_2 \nonumber \\
&& +(p_1^{st}k_{1}+ p_2^{st}k_{2})\nu, 
\end{eqnarray}
and $\nu=\alpha+\beta$, $k_1=k_{\rm f1}+k_{\rm b1}$,  
$k_2=k_{\rm f2}+k_{\rm b2}$.
In Eq. (\ref{pd}), $\langle \tilde p_{D}(s)\rangle$ 
denotes the Laplace transform,  $\langle \tilde p_{D}(s)\rangle=
\int_0^{\infty}
\exp(-s t) \langle p_{D}(t)\rangle dt$. The averaged asymptotic population of 
the donor level follows as $\langle p_{D}(\infty)\rangle=
\tilde A(0)/\tilde B(0)$,
\begin{eqnarray}\label{asymp}
\langle p_{D}(\infty)\rangle=
\frac{p_2^{st}k_{\rm b2}(k_1+\nu)+p_1^{st}k_{\rm b1}(k_2+\nu)}
{k_1k_2+(p_1^{st}k_{1}+p_2^{st}k_{2})\nu}.
\end{eqnarray}  
The averaged relaxation of the donor state population is obtained 
by the inversion
of Eq. (\ref{pd}) to the time domain. It is bi-exponential and reads
\begin{eqnarray}\label{pd-time}
\langle  p_{D}(t)\rangle=\langle p_{D}(\infty)\rangle+[1-\langle
p_{D}(\infty)\rangle]R(t),
\end{eqnarray}
where $R(t)=\sum_{i=1,2}c_i\exp(-\Gamma_i t)$ is the relaxation function
with the rate constants
\begin{eqnarray}\label{rates}
\Gamma_{1,2}&= &\frac{1}{2}\Big[k_1+k_2+\nu \nonumber  \\ 
&\pm &  \sqrt{(k_1+\alpha-k_2-\beta)^2+
4\alpha\beta}\Big]
\end{eqnarray}
 and the weighting coefficients
 \begin{eqnarray}\label{c12}
c_{1,2}=\frac{1}{2}\left [1\pm \frac{k_1+k_2+\alpha+\beta-2c_0/a_0}
{\sqrt{(k_1+\alpha-k_2-\beta)^2+
4\alpha\beta}}   \right].
\end{eqnarray}
The remaining quantities $a_0$ and $c_0$ in Eq. (\ref{c12}) are
\begin{eqnarray}\label{ca}
a_0& =&\alpha k_{\rm f2}(k_1+\alpha)+\beta k_{\rm f1}(k_2+\beta)+\alpha\beta(k_{\rm
f1}+k_{\rm f2}), \nonumber \\
c_0& = &\nu^2(\alpha k_{\rm f2}+\beta k_{\rm f1})+
\alpha k_1 k_{\rm f2}(2\alpha+k_1) \nonumber \\
&+&\beta k_2 k_{\rm f1}(2\beta+k_2)
+\alpha\beta (k_1+k_2)(k_{\rm f1}+k_{\rm f2}).
\end{eqnarray}  
The distribution of the first arrival times at the acceptor state 
$\psi_D(\tau)$ can be immediately obtained from the survival probability
$\Phi_D(\tau)$ which in turn follows from the above
relaxation function $R(\tau)$ by setting $ k_{\rm b1},k_{\rm b2}\to 0$, i.e.
by assuming that the acceptor state is absorbing. This yield
immediately
\begin{eqnarray}\label{survival}
\Phi_D(\tau)=\sum_{i=1,2}c_i\exp(-\Gamma_i\tau),
\end{eqnarray}
where the rate constants $\Gamma_i$ and the coefficients $c_i$ reduce
to
\begin{eqnarray}\label{rates-f}
\Gamma_{1,2}& = &\frac{1}{2}\Big[k_{\rm f1}+k_{\rm f2}+\nu \nonumber \\
&\pm & \sqrt{(k_{\rm
f1}+\alpha-k_{\rm f2}-\beta)^2+
4\alpha\beta}\Big]
\end{eqnarray} 
and 
 \begin{eqnarray}\label{c12-f}
c_{1,2}=\frac{1}{2}\left [1\mp \frac{\alpha+\beta+(k_{\rm f2}-k_{\rm
f1})(\beta-\alpha)/(\alpha+\beta)}
{\sqrt{(k_{\rm f1}+\alpha-k_{\rm f2}-\beta)^2+
4\alpha\beta}} \right],
\end{eqnarray}
respectively. These are the same expressions as, e.g., Eqs. (7)-(10) 
in Ref. \cite{JCP05}
obtained there using a different method.
The corresponding mean forward transfer time is
\begin{eqnarray}\label{tau}
\langle \tau_f\rangle =\frac{(\alpha+\beta)^2+\alpha k_{\rm f1}+\beta k_{\rm
f2}}{(\alpha+\beta)[\alpha k_{\rm f2}+\beta k_{\rm
f1}+k_{\rm f1}k_{\rm f2}]}.
\end{eqnarray}
All the quantities, we are interested in, are thus formally determined.

\subsection{Conditions for pumping}

Let us suppose that the ET is characterised by Eq. (\ref{Marcus})
with the negative bias $\epsilon_1<0$ and the tunnelling matrix element $V_{\rm tun1}$
in conformation 1 and the positive bias $\epsilon_2>0$ and the 
tunnelling matrix element $V_{\rm tun2}$ in conformation 2. In addition, one
assumes that the reorganisation energy $\lambda$ is the same in both conformations.
Then  for the ratio of the averaged donor and acceptor populations we obtain
from Eq. (\ref{asymp}):
\begin{eqnarray}\label{ratio}
\frac{\langle p_D(\infty)\rangle}{\langle p_A(\infty)\rangle}=
\exp\left(-\frac{\epsilon_2}{k_BT}\right)\frac{1+\zeta\xi}{1+\zeta/\xi}\,,
\end{eqnarray} 
where 
\begin{eqnarray}\label{zeta}
\zeta=\frac{p_1^{st}}{p_2^{st}}\frac{(\nu+k_2)F\cosh(\epsilon_2/2k_BT)}
{\nu\cosh(\epsilon_2/2k_BT)+k_2F\cosh(\epsilon_1/2k_BT)}, \\
F=\left( \frac{V_{\rm tun1}}{V_{\rm tun2}}\right)^2 \label{F}
\exp[-(\epsilon_1^2-\epsilon_2^2)/(4\lambda k_BT)]
\end{eqnarray}
and $\xi=\exp[-(\epsilon_1-\epsilon_2)/(2k_BT)]$.  The pumping
 is most efficient when 
$\langle p_D(\infty)\rangle/\langle p_A(\infty)\rangle\ll 1$. 
This requires $\epsilon_2\gg k_BT$ and $\zeta\xi\ll 1$. Moreover, 
to have the averaged energy gained by the transferring electrons maximal, i.e.
 $\overline \epsilon =p_1^{st}\epsilon_1+p_2^{st}\epsilon_2\approx 
 \epsilon_1$, one has to ensure that $p_2^{st}\ll p_1^{st}$ which 
 contradicts, however, at the first look 
 to the condition $\zeta\ll 1$ in Eq. (\ref{zeta}).
 To resolve this contradiction, one requires sufficiently
 small values of $F\ll 1$ and this in turn demands 
 $V_{\rm tun1}\ll V_{\rm tun2}$. How small is small depends on 
 the system parameters.    
For example, for nitrogenase 
 $\Delta \epsilon=\epsilon_2-\epsilon_1\approx 200$ meV \cite{Kurnikov}.
 Therefore, at the room temperatures, $k_BT\approx 25$ meV, 
 $\xi\approx 54.6$ is pretty large.
Furthermore, let us assume for simplicity that $|\epsilon_1|=\epsilon_2$, 
so that 
  $\zeta=(\beta/\alpha)(\nu+k_2)F/(\nu+k_2F)$ 
  and $F=(V_{\rm tun1}/V_{\rm tun2})^2$. Moreover,
  we assume for a moment that 
  $\alpha=0.1\beta$, so that the conformation 1 is about ten times more
  probable than the conformation 2. Then, to satisfy $\zeta\xi\ll 1$ and to 
  have an efficient pumping, 
  $V_{\rm tun1}$ should be smaller than $V_{\rm tun2}$ by, at least,
  two orders of magnitude. In such a case,  $k_2\gg k_1$, and for
  $k_1\ll \nu\ll k_2$, one  can expect that the 
  overall transfer will become
  gated by the conformation fluctuations.  Namely, it follows from Eq.
  (\ref{tau}) that for $k_{\rm f1}\ll \alpha\ll \beta\ll k_{\rm f2}$,
  $\langle \tau_f\rangle \approx 1/\alpha$ \cite{JCP05}, i.e. the pumping
  of electrons is locked to the conformational transitions caused by
  the binding of ATP molecules somewhere
   nearby the electron donor site (scheme I)\cite{foot3}, or vice
  versa by their hydrolysis and dissociation nearby the acceptor
  site (in the scheme II). 

\section{Results and discussion}
  
\begin{figure}[ht]
\begin{center}
 \vspace{0.5cm}
 \includegraphics[width=8cm]{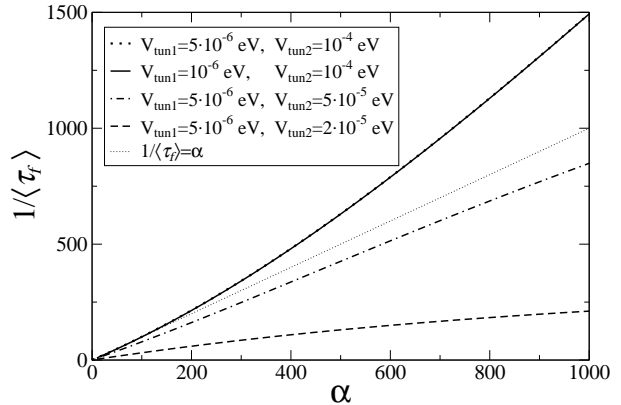}
 \end{center}
 \caption{Dependence of the inverse of mean forward transfer time
 on the rate of conformational transitions  $\alpha \propto \rm [ATP]$
 (scheme I is assumed).
 \label{Fig3}}
  \end{figure}

\begin{figure}[ht]
\begin{center}
 \vspace{0.5cm}
 \includegraphics[width=8cm]{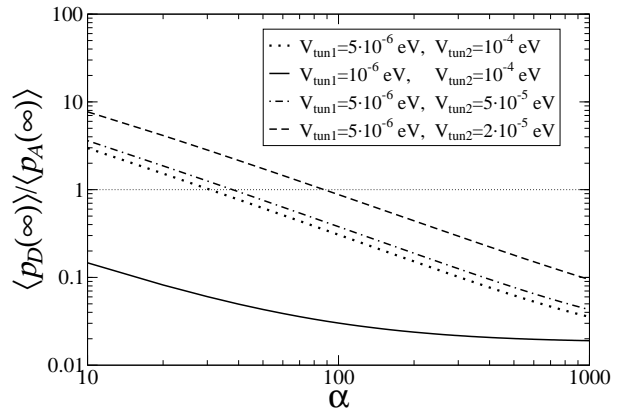}
 \end{center}
 \caption{Dependence of the ratio of donor and acceptor populations
 on the rate of conformational transitions  $\alpha \propto 
 \rm [ATP]$.\label{Fig4}}
  \end{figure}

  The outlined mechanism should be very robust. 
  We illustrate it in Figs. \ref{Fig3}, \ref{Fig4}  
  for the following  realistic test parameters which
partially correspond to nitrogenase \cite{Kurnikov} and partially
are chosen just to demonstrate the essential effects:
$\lambda=1.2$ eV, $\epsilon_1=-0.1$ eV, $\epsilon_2=0.1$ eV;
the tunnelling coupling energies are given in the Table \ref{Table}
together with the corresponding Marcus rates.
Furthermore, the value 
 $\beta=1000$ $\rm sec^{-1}$
  is used in calculations and the rate $\alpha$ is varying as a control
  parameter assuming its proportionality to the ATP concentration [ATP].

\begin{table}
  \caption{Tunnelling coupling energies (in eV) and the 
  corresponding Marcus rates (in $\rm sec^{-1}$)}
{\begin{tabular}{@{}lcccc}\toprule
   Tunnelling coupling energies
  & $k_{\rm f1}$ & $k_{\rm b1}$ & $k_{\rm f2}$ & $k_{\rm b2}$ \\
\colrule
$V_{\rm tun1}=5\cdot 10^{-6}$, $V_{\rm tun2}=10^{-4}$ & 0.54 
&   29.5   & $1.18\cdot 10^4$ & 216 \\
$V_{\rm tun1}=10^{-6}$, $V_{\rm tun2}=10^{-4}$   
& 0.02
&   1.18  & $1.18\cdot 10^4$ & 216 \\
$V_{\rm tun1}=5\cdot 10^{-6}$, $V_{\rm tun2}=5\cdot 10^{-5}$ & 0.54 
&   29.5   & $2.95\cdot 10^3$ & 53.9 \\
$V_{\rm tun1}=5\cdot 10^{-6}$, $V_{\rm tun2}=2\cdot 10^{-5}$ & 0.54 
&   29.5   & $471$ & 8.63 \\
   \botrule
  \end{tabular}}
\label{Table}
\end{table}

As it is clearly seen in Figs. \ref{Fig3}, \ref{Fig4}
the pumping effect is indeed present for 
$V_{\rm tun2}/V_{\rm tun1}=100$ (continuous line). 
Moreover, the mean forward time is locked to the rate $\alpha$
for $\alpha<200$ $\rm sec^{-1}$, being practically its inverse. Moreover,
the calculation of the survival probability $\Phi_D(t)$, e.g.
for $\alpha=100$ $\rm sec^{-1}$, $\Phi_D(t)\approx 
0.08\exp(-12790\,t)+0.92\exp(-92\,t)$; $\alpha=200$ $\rm sec^{-1}$,
$\Phi_D(t)\approx 0.14\exp(-12798\,t)+0.86\exp(-184\,t)$, etc.,
shows that the transfer is almost single exponential and
$\langle\tau_f\rangle^{-1}\approx \alpha$ can be regarded as 
the pumping rate. With the further increase of $\alpha$, the transfer
kinetics becomes, however, ever more nonexponential with the effective 
rate $\langle\tau_f\rangle^{-1}$ being larger than $\alpha$, cf. 
Fig. \ref{Fig3}. Here, the conformational transitions cease gradually
 to be the rate-limiting step and the tunnelling time $1/k_{\rm f2}$
becomes ever more important for the overall kinetics. 
However, this regime presents
lesser interest in the present context since 
$\overline \epsilon=0$ for $\alpha=\beta$ and the pumping effect then 
vanishes. 

Furthermore, let us to keep $V_{\rm tun2}$ the same, but to 
increase $V_{\rm tun1}$ such that the ratio $V_{\rm tun2}/V_{\rm tun1}$
becomes much smaller, $V_{\rm tun2}/V_{\rm tun1}=20$ (fat dotted
lines in Figs. \ref{Fig3}, \ref{Fig4}). This does not affect 
$\langle\tau_f\rangle$ in Fig. \ref{Fig3}, but changes dramatically
the ratio of populations in Fig. \ref{Fig4}. The pumping effect is present
for sufficiently large rates $\alpha$ (cf. negative values of 
$\langle p_D(\infty)\rangle/\langle p_A(\infty)\rangle$ in Fig. \ref{Fig4}). 
The pumping efficiency drops, however,
essentially.  Moreover, the critical
values of the rate $\alpha_c$ and the associated ATP 
concentration $\rm [ATP_c]$ emerge.
The overall transfer occurs in the ``donor$\to$acceptor'' direction
if only $\alpha>\alpha_c$. On the other hand, one must keep $\alpha<\beta$.
Otherwise, the transferred electrons will start to lose in energy on average.
Clearly, such a pump would not function perfectly. 
To realize a good electronic pump, the ratio 
$V_{\rm tun2}/V_{\rm tun1}$ must be large.

\begin{figure}[ht]
\begin{center}
 \vspace{0.6cm}
 \includegraphics[width=8cm]{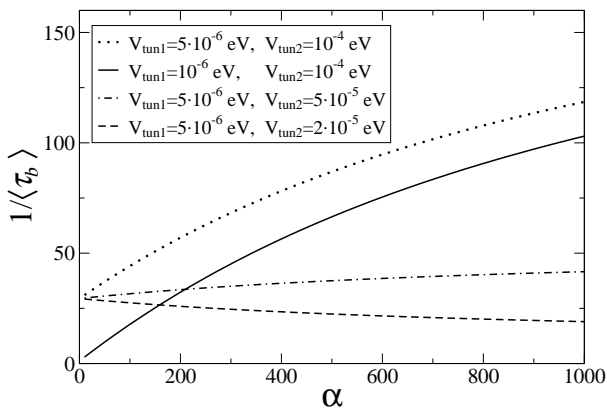}
 \end{center}
 \caption{Dependence of the inverse of mean backward transfer time
 on the rate of conformational transitions  $\alpha$.\label{Fig5}}
  \end{figure}

The inversion of the transfer direction
depending on the rate of ``1 $\to$ 2'' 
conformational transition, cf. Fig. \ref{Fig4} at several combinations
of the tunnelling couplings, is rather intriguing.  Namely, for 
sufficiently small $\alpha$ 
the roles of the donor 
and acceptor states are interchanged. Here, the effective rate of the backward
``acceptor$\to$donor''  ET, 
defined as the inverse of the corresponding mean first
passage time $\langle \tau_b\rangle$, can be used to quantify the rate of 
transfer in this direction (assuming $\langle p_D(\infty)\rangle/
\langle p_A(\infty)\rangle\gg 1$).
$\langle \tau_b\rangle$ can be obtained from Eq. (\ref{tau}) by setting there
$k_{\rm f1}\to k_{\rm b1}, k_{\rm f2}\to k_{\rm b2}$. This quantity is depicted
in Fig. \ref{Fig5}.
It is clearly seen in Fig. \ref{Fig5} that the effective backward rate is not
gated by the conformational fluctuations for the used sets of parameters.
Nevertheless, it can strongly depend on $\alpha$,  being almost 
linearly proportional
to $\alpha$ at small $\alpha$ (see the continuous line in Fig. \ref{Fig5}). 
It might thus
resemble  a gating regime. 

One more interesting feature is that with the decrease of 
$V_{\rm tun2}$ the effective transfer rate becomes smaller than $\alpha$,
with the tunnelling providing the rate-limiting step when
$\alpha$ increases (see dashed line in Fig. \ref{Fig3}).

\subsection{Pumping efficiency}

The maximal pumping efficiency can be  
defined as the averaged energy gained by the 
transferred electron relative to the energy required
to drive one transfer cycle, i.e.  
$\eta=|\overline{\epsilon}|/
\Delta G_{\rm drive}$, or
\begin{eqnarray}
\eta=
\frac{p_1^{st} |\epsilon_1|-p_2^{st} \epsilon_2}{\epsilon_2+|\epsilon_1|} \; .
\end{eqnarray}
For the above parameters (corresponding to the continuous lines in Figs. \ref{Fig3},
\ref{Fig4} and $\alpha\sim 100$), the maximal 
pumping efficiency 
is rather high approaching $\eta=0.5$  \cite{foot4}. It can be even 
higher approaching one, 
if the affinity
of the donor state D1 to electrons is much larger than the affinity of
the state D2 (scheme I), i.e. the protein complex takes preferably
 electrons from the bulk in the conformation I (the formal 
 solution of the model has to be modified in this case, but the qualitative
 features remain).

\section{Conclusions}

The considered generic model might seem somewhat oversimplified. It is indeed 
aimed primarily
to highlight the basic working principles and their practical relevance. 
This model should be extended and generalised further 
in several directions, e.g. a correlated 
two-electron transfer should probably be considered in nitrogenase as an 
elementary step rather than single-electron transfer and a proper 
treatment of the ATP binding, 
hydrolysis and dissociation of the hydrolysis products would require to introduce 
more conformations than two. Moreover, the external uptake and release of
electrons from and to the donor and acceptor sites, e.g., from mobile electron 
carriers should be incorporated
in the complete model. Nevertheless, the considered elementary model 
does allow to manifest the main operating principles which are not 
much different
from those well established and clearly understood, both phenomenologically
and in progressing details,  
for ionic pumps \cite{Alberts,NossalLecar,KeenerSneyd}. Moreover,
it allows one to clarify some important conditions for the efficient
pumping such as a large
ratio of the tunnelling couplings in the different conformations  and
a possible existence of the critical ATP concentrations.    
The profound physical difference between the ionic and electronic
pumps is, however, 
that the electron is essentially a quantum particle
and it tunnels over a large distance between metalloclusters in 
nitrogenase (also in other electron transfer complexes, like cytochrome $bc_1$) 
using virtually protein bridging states. This is why the {\it details} here
are definitely very different from ionic pumps. 
They do matter and are important to arrive
in a future at the detailed (quantum)-mechanistic, molecular-dynamic 
understanding which still is lacking at present. Unlike to the many-years, extensive
research on ionic pumps we undertake here really the first steps.
The research domain of electron transfer driven by a chemical energy
source through nonequilibrium 
conformational fluctuations is just emerging.


\begin{thebibliography}{100}
\bibitem{foot1}In 
the living cells ATP is perpetually 
synthesised from ADP and $\rm P_i$ by the ATP-syntase membrane 
complexes fuelled by 
the electrochemical proton gradient
\cite{Alberts,Nelson}.
\bibitem{foot2}
Non-Markovian generalisation of this scheme accounting for 
complex dynamics with memory within a conformation 
can be done, e.g., in the framework of  a
stochastic trajectories description, cf.
Refs. \cite{Kampen79,PRE04,JCP05,AdvPhys05}.
\bibitem{foot3}``Nearby'' can be
   actually $15\; \rm \AA$ apart, with the acceptor site being
   about $30\;\rm \AA$ apart and buried
   inside of protein, like in the case of nitrogenase.
\bibitem{foot4} 
It compares well with the efficiency
of various ionic pumps which can be as high as 
$\eta=0.75$ \cite{Nelson}.
\end{thebibliography}
 \end{document}